\documentclass[aps,prb,groupedaddress,reprint,superscriptaddress]{revtex4-1}
\usepackage{amsfonts}
\usepackage{amsmath}\usepackage{amssymb}
\usepackage{graphicx}
\usepackage{dcolumn}
\usepackage{bm}
\usepackage{color}
\usepackage[normalem]{ulem}
\begin{document}

\title{Isotope effects in x-ray absorption spectra of liquid water}

\author{Chunyi Zhang}
\affiliation{Department of Physics, Temple University, Philadelphia, Pennsylvania 19122, USA}

\author{Linfeng Zhang}
\affiliation{Program in Applied and Computational Mathematics, Princeton University, Princeton, New Jersey 08544, USA}	

\author{Jianhang Xu}
\affiliation{Department of Physics, Temple University, Philadelphia, Pennsylvania 19122, USA}

\author{Fujie Tang}
\affiliation{Department of Physics, Temple University, Philadelphia, Pennsylvania 19122, USA}

\author{Biswajit Santra}
\affiliation{Department of Physics, Temple University, Philadelphia, Pennsylvania 19122, USA}

\author{Xifan Wu}
\email{xifanwu@temple.edu}
\affiliation{Department of Physics, Temple University, Philadelphia, Pennsylvania 19122, USA}


\begin{abstract}
The isotope effects in x-ray absorption spectra of liquid water are studied by a many-body approach within electron-hole excitation theory. The molecular structures of both light and heavy water are modeled by path-integral molecular dynamics based on the advanced deep-learning technique. The neural network is trained on \textit{ab initio} data obtained with SCAN density functional theory. The experimentally observed isotope effect in x-ray absorption spectra is reproduced semiquantitatively in theory. Compared to the spectrum in normal water, the blueshifted and less pronounced pre- and main-edge in heavy water reflect that the heavy water is more structured at short- and intermediate-range of the hydrogen-bond network. In contrast, the isotope effect on the spectrum is negligible at post-edge, which is consistent with the identical long-range ordering in both liquids as observed in the diffraction experiment.

\end{abstract}


\maketitle

\section{INTRODUCTION}

Water is one of the most important substances to make life possible on earth \cite{franks_water_2007, ball_water_2008, bellissent-funel_water_2016}.
The unique hydrogen (H)-bond network results in the distinctive properties of water and has been at the center of scientific interest for decades \cite{stillinger_water_1980, luzar_hydrogen-bond_1996, fecko_ultrafast_2003, pettersson_watermost_2016, nilsson_structural_2015}.
Normal water (H$_2$O) and heavy water (D$_2$O) only differ slightly in the H-bond network \cite{soper_quantum_2008, zeidler_isotope_2012}; however, the former is essential for a living cell, while the latter can be harmful in many ways \cite{franks_water_2007, thomson_physiological_2006, kushner_pharmacological_1999}. Moreover, minute distortions in the H-bond network can cause noticeable changes in functionalities of numerous biological processes occurring in aqueous environments \cite{fersht_hydrogen_1985, ohtaki_structure_1993, leberman_effect_1995, tarek_role_2002, chen_hydroxide_2018}. Therefore, a precise picture of the isotope effect of liquid water is crucial, which also serves as an important milestone to accurately understand the intensely discussed microscopic structure of water \cite{wernet_structure_2004, nilsson_x-ray_2010, tse_x-ray_2008, smith_probing_2006, nilsson_structural_2015, fransson_x-ray_2016, amann-winkel_x-ray_2016, smith_soft_2017}.

The last decade has witnessed a rapid emergence of the x-ray absorption spectroscopy (XAS) being applied to probing the H-bond network of water \cite{wernet_structure_2004, nilsson_x-ray_2010, tse_x-ray_2008, smith_probing_2006, nilsson_structural_2015, fransson_x-ray_2016, amann-winkel_x-ray_2016, smith_soft_2017}. In the XAS process, the time scale of the electron-hole interaction is much shorter than that of the molecular relaxation \cite{bernath_spectra_2015}. Therefore, XAS carries out an instantaneous local fingerprint of water structure, which is complementary to the averaged structural information obtained in diffraction experiments \cite{soper_quantum_2008, zeidler_oxygen_2011, zeidler_isotope_2012, soper_is_2019}. Light and heavy water have been extensively studied by various experimental techniques, however their differences in XAS only became available very recently by the increased spectral resolution in the transmission-mode spectroscopy technique \cite{schreck_isotope_2016}. It revealed that they are similar but not identical. The discernible spectral difference suggests H-bond networks in H$_2$O and D$_2$O are affected differently by nuclear quantum effects (NQEs) \cite{schreck_isotope_2016}.

With the significant advances in XAS experiment, the theoretical exploration of XAS spectra is urgently needed to unambiguously associate spectral features to specific structural motifs of water, which requires both an accurate molecular structure and a proper treatment of electron-hole interaction.
Based on density functional theory (DFT) \cite{hohenberg_inhomogeneous_1964}, Feynman path-integral \textit{ab initio} molecular dynamics (PI-AIMD) \cite{car_unified_1985, marx_ab_1996, ceriotti_nuclear_2013, ceriotti_i-pi:_2014, ceriotti_nuclear_2016} provide an ideal platform to predict the liquid structure by including the NQEs. However, for decades, simulation of water has been a formidable task. Extensive studies have identified that the van der Waals interaction and exact exchange are key ingredients to differentiate between water and ice \cite{wernet_structure_2004, zhang_structural_2011, wang_density_2011, mogelhoj_ab_2011, distasio_individual_2014, gaiduk_density_2015, miceli_isobaric_2015, del_ben_probing_2015, chen_ab_2017}. To accommodate these fine effects, a non-local exchange-correlation functional should be adopted in functional construction that requires higher rungs \cite{sun_strongly_2015, perdew_rationale_1996, adamo_toward_1999} in the metaphorical Jacob's ladder \cite{perdew_jacobs_2001}. In this regard, the modeling of water by SCAN functional \cite{sun_strongly_2015} has shown great accuracy in comparison to experiment \cite{chen_ab_2017}. In parallel, the modeling of electron excitation in the optical process stands as another major challenge that has been under active development for years \cite{hybertsen_electron_1986, hetenyi_calculation_2004, prendergast_x-ray_2006, kang_enhanced_2010, rohlfing_electron-hole_2000, chen_x-ray_2010, kong_roles_2012, sun_x-ray_2017, sun_electron-hole_2018}. The excited electrons need to be treated as quasiparticles to solve the Bethe-Salpeter equation (BSE) \cite{vinson_bethe-salpeter_2011, fransson_x-ray_2016}, whose Coulomb interactions are screened by the electron sea in water. The proper treatments of the electronic screening, such as the Slater's transition state theory \cite{slater_nonintegral_1969, slater_statistical_1970} or the more rigorous Hedin's GW approximation \cite{hedin_new_1965, hedin_effects_1970} for the self-energy approach, is found to be crucial to qualitatively reproducing experimental XAS spectra. However, due to the significantly increased computational burden in solving BSE as well as in the PI-AIMD simulation, such theoretical studies so far remain elusive.

To address the above issues, we compute the XAS spectra of both H$_2$O and D$_2$O at oxygen $K$ edge based on the self-energy approximation to the BSE. In particular, the liquid structures are generated from path-integral deep potential molecular dynamics (PI-DPMD) using a deep neural network-based potential energy model \cite{wang_deepmd-kit:_2018, zhang_deep_2018, zhang_active_2019, zhang2018end, ko_isotope_2019}.
The PI-DPMD scheme preserves the accuracy of SCAN-DFT with a computational cost comparable to that of empirical force fields.
The resulting isotope effects in XAS spectra are in good agreement with experiment \cite{schreck_isotope_2016}, which shows a stronger influence by NQEs in light water than in heavy water.
The pre-edge of the XAS spectra, a signature of short-range ordering of H-bond network, shows a blueshift in the excitation energies and weaker spectral intensities in D$_2$O compared to H$_2$O, which originates from the shorter covalent bond but a stronger H-bonding environment in D$_2$O. For intermediate-range ordering, the light water exhibits an enhanced degree of inhomogeneity as revealed by local structure index analysis \cite{shiratani_growth_1996, duboue-dijon_characterization_2015, santra_local_2015}. Therefore, a more pronounced main-edge at lower energy is identified in H$_2$O because a softer liquid structure promotes the localization and stabilization of the excitons. The post-edge of XAS as an indicator of long-range ordering, however, has a negligible isotope effect. This is consistent with the nearly identical structures beyond second shell coordination as observed in the diffraction experiment \cite{soper_quantum_2008}. This work simulates the isotopic differences of XAS spectra of liquid water. Our approach, combining accurate molecular dynamics simulations and the electron-hole theory, provides an important theoretical lens to understand the fine structures and quantum fluctuations of water by XAS.

\section{METHOD}

The PI-DPMD simulations were conducted in an isobaric-isothermal ensemble at 330 K and 1 bar for 0.3 ns with a 128-molecule supercell. For both H$_2$O and D$_2$O, one representative snapshot was selected from PI-DPMD trajectories and was adopted for the calculation of XAS spectra using our recently developed enhanced static Coulomb-hole and screened exchange approximation \cite{kang_enhanced_2010, sun_x-ray_2017}. The XAS spectra of H$_2$O and D$_2$O were aligned according to the position of the post-edge because the post-edges of H$_2$O and D$_2$O coincide in experiment \cite{schreck_isotope_2016}. More simulation details are described in the Supplemental Material \cite{support}.

\section{RESULTS AND DISCUSSION}

\begin{figure}
	\includegraphics[width=0.47\textwidth]{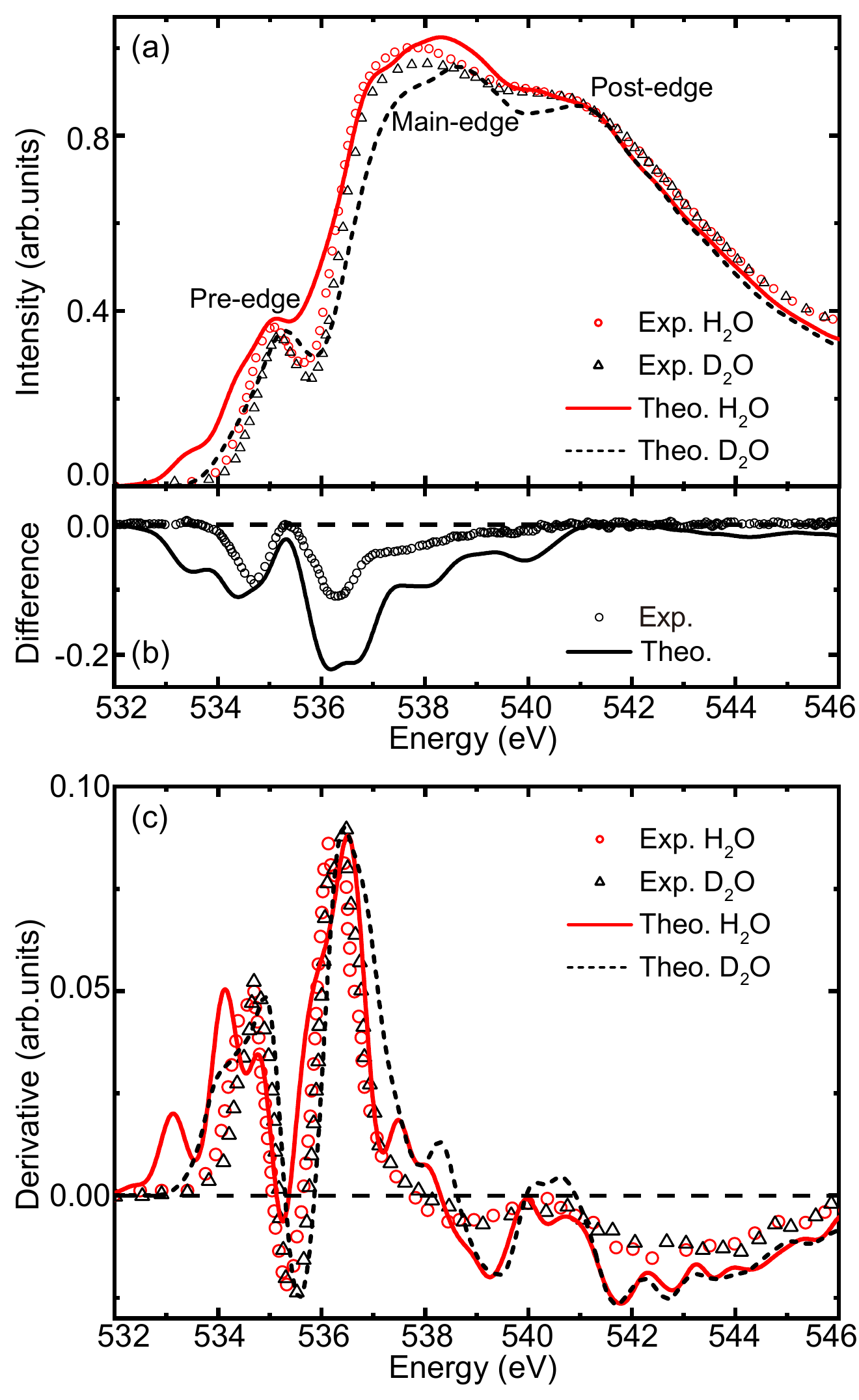}
	\caption{(a) XAS spectra of H$_2$O (red) and D$_2$O (black) from theory (solid and dashed lines) and experiment \cite{schreck_isotope_2016} (circles and triangles). (b) D$_2$O-H$_2$O difference of the XAS spectra from theory (line) and experiment \cite{schreck_isotope_2016} (circles). (c) First-order derivatives of the XAS spectra with respect to the energy of H$_2$O (red) and D$_2$O (black) from theory (solid and dashed lines) and experiment (circles and triangles).}\label{fig:xas}
\end{figure}

The theoretical XAS spectra of both liquids H$_2$O and D$_2$O are presented in Fig.~\ref{fig:xas}(a) together with the experimental spectra \cite{schreck_isotope_2016}.
A good agreement can be seen between the theory and experiment, not only on overall spectral shapes and all three features of pre-edge ($\sim$535 eV), main-edge ($\sim$538 eV), and post-edge ($\sim$541 eV) in Fig.~\ref{fig:xas}(a), but also on the more delicate spectral differences between the two isotopes in Fig.~\ref{fig:xas}(b). (The small peak at $\sim$ 533.5 eV in the theoretical results in Fig.~\ref{fig:xas} is caused by nuclear quantum fluctuation, which can be eliminated by averaging over more snapshots.)
Within the same excitation energy scale, the spectrum of light water is slightly broader than that of the heavy water. In particular, the blueshifts of the peaks of pre- and main-edge are about 160 and 320 meV, which are close to the experimental values of 120$\pm$20 and 200$\pm$20 meV \cite{schreck_isotope_2016}, respectively. A close inspection of the spectral difference between H$_2$O and D$_2$O in Fig.~\ref{fig:xas}(b) further reveals that the isotope effect in XAS is most significant in the pre- and main-edges, which decays rapidly and is negligible at the higher excitation energies in the post-edge. Apart from the XAS spectra and difference spectra, the first-order derivatives of the XAS spectra obtained in this work also agree well with experimental results \cite{schreck_isotope_2016} as shown in Fig.~\ref{fig:xas}(c).
The successful prediction of the experimental measurement indicates that the H-bond structures and their signatures in electronic excitations of H$_2$O and D$_2$O are both accurately modeled.

\begin{figure}
	\includegraphics[width=0.48\textwidth]{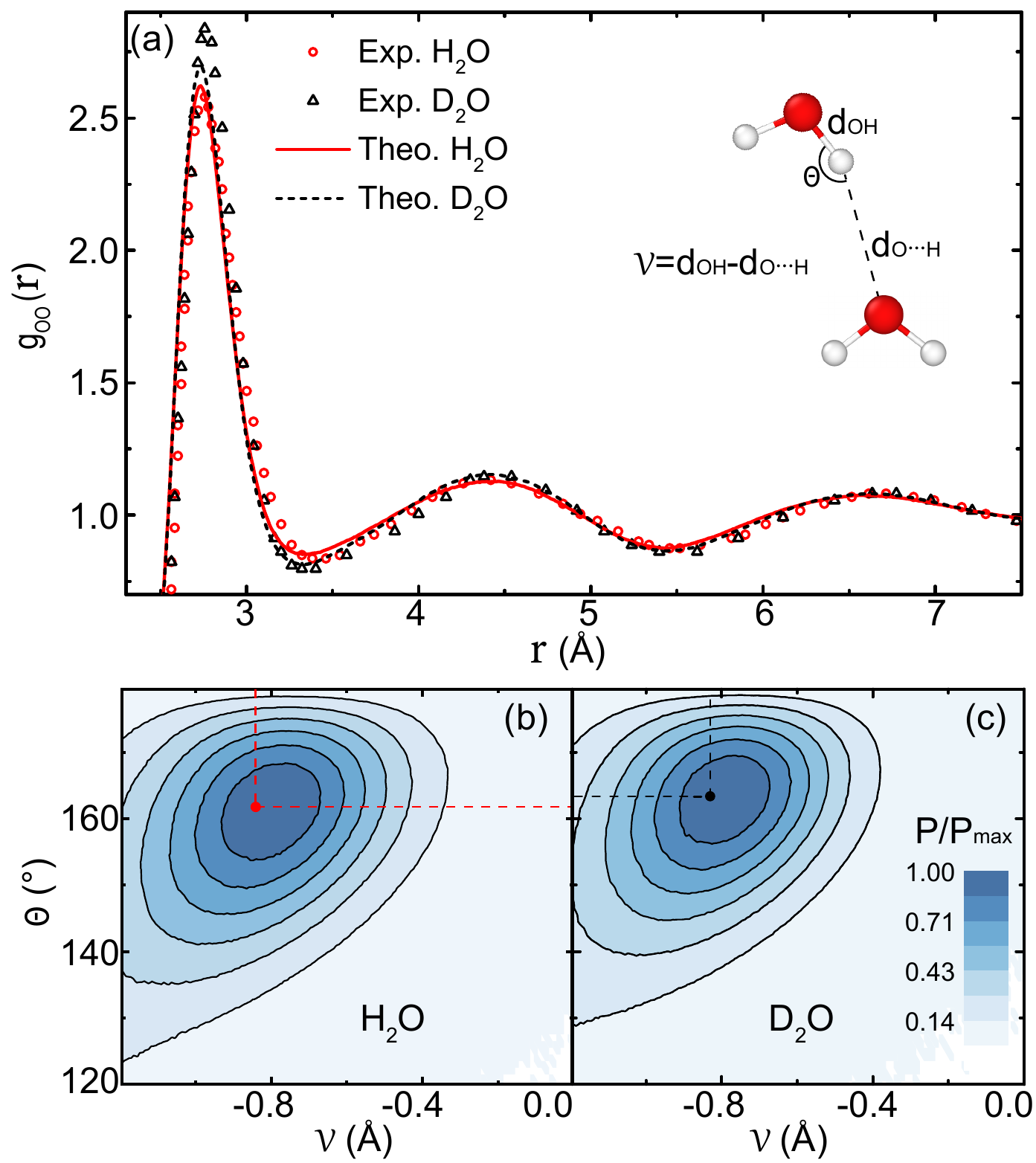}
	\caption{(a) ${g\rm_{OO}}(r)$ of liquids H$_2$O (red) and D$_2$O (black) from theory (solid and dashed lines) and experiment \cite{soper_quantum_2008} (circles and triangles). The inset shows definitions of proton transfer coordinate $\nu$ and OH$\cdots$O angle $\theta$. Joint probability distributions of $\theta$ as a function of $\nu$ for (b) H$_2$O and (c) D$_2$O. The red dot located at ($-$0.84, 161.80) in (b) and the black dot located at ($-$0.83, 163.34) in (c) show the $\theta$ and $\nu$ values with the highest probability.}\label{fig:gr}
\end{figure}

In experiment, heavy water is characterized as a more structured liquid than light water \cite{soper_quantum_2008}. The subtle structural differences are accurately predicted as displayed in Fig.~\ref{fig:gr}(a). The oxygen-oxygen pair distribution functions, ${g\rm_{OO}}(r)$, show more prominent first and second coordination shells in D$_2$O than those in H$_2$O, which agrees well with the diffraction measurement by Soper {\it{et al.}} \cite{soper_quantum_2008}. Since both heavy and light water share the same electronic configuration, their structural difference arises entirely from the NQEs that restructure their H-bond networks by different magnitudes. Under the influence of NQEs, the protons are more delocalized and are able to probe the configuration space that is inaccessible to the classical nuclei.
The delocalized protons along the direction of the stretching mode promote the formation of H-bonds. As shown by the proton transfer coordinate \cite{wang_quantum_2014} in Fig. 2, the tendency of a proton to approach the acceptor molecule is increased under the larger NQEs in H$_2$O than in D$_2$O.
On the contrary, the delocalized proton along the direction of the libration mode facilitates the breaking of H-bonds by perturbing the OH$\cdots$O angle, $\theta$, further away from 180$^\circ$. The overall NQEs resulting from the two competing effects are dependent on the anharmonicity of the potential energy surface \cite{li_quantum_2011, ceriotti_nuclear_2016}. In liquid water, NQEs actually soften the liquid structure with more broken H-bonds \cite{morrone_nuclear_2008, cheng_ab_2019}.
Because of the heavier deuteron than proton, NQEs are suppressed in heavy water, which can be identified by the narrower distribution of deuteron as functions of $\nu$ and $\theta$ comparing to that in light water as shown in Fig.~\ref{fig:gr}(b) and (c).
Not surprisingly, the H-bond network in D$_2$O is less destructed by the NQEs compared to H$_2$O. The slightly stronger H-bond of D$_2$O can be seen by the more parallelly aligned OH$\cdots$O angle and a proton transfer coordinate that is closer to zero than those of H$_2$O as shown by the red and black dots in Figs.~\ref{fig:gr} (b) and \ref{fig:gr} (c). The less perturbed H-bond network of D$_2$O than H$_2$O by NQEs is captured by distinct edge features in XAS spectra.

\begin{figure*}
	\includegraphics[width=1.0\textwidth]{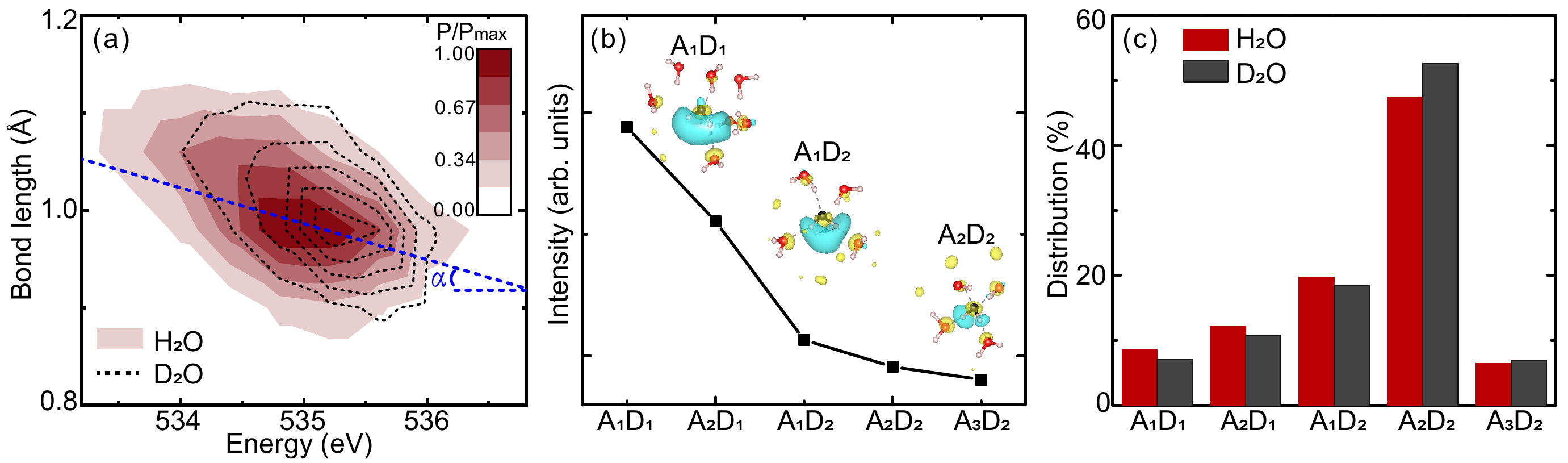}
	\caption{(a) Joint probability distribution of O-H/O-D covalent bond lengths as a function of excitation energies of states with $4a_1$ character of liquids H$_2$O (red shadows) and D$_2$O (dashed black lines).
(b) Averaged intensities of states with $4a_1$ character as a function of A$_i$D$_j$ in H$_2$O. A$_i$D$_j$ indicates the number of acceptor-type (A$_i$) and donor-type (D$_j$) H-bonds, respectively. Insets show representative H-bond environments and distributions of quasiparticle wavefunctions (QWs). The excited oxygen atom is shown in black. QWs with opposite signs are depicted in blue and yellow.
(c) Distributions of A$_i$D$_j$ in liquids H$_2$O (red) and D$_2$O (grey). }\label{fig:pre}
\end{figure*}

The pre-edge is attributed to a bound exciton with $a_1$ characteristic whose origin can be traced back to the first electronic excitation in water vapor \cite{chen_x-ray_2010, sun_x-ray_2017}. Once the water molecule is excited, a positive oxygen core-hole is left behind. The core-hole generates a strong potential that traps excitonic states that are well localized within the molecule. Therefore, the pre-edge carries out a signature of short-range ordering of the H-bond network \cite{chen_x-ray_2010, sun_x-ray_2017}. The excitation energies are sensitive to the relative position between proton and oxygen as determined by the covalent bond length, which is under constant thermal and quantum fluctuations.
When the bond length becomes longer, the proton moves away from the excited oxygen. As a result, the enhanced electropositivity around the oxygen atom makes the core-hole Coulomb potential effectively stronger, which stabilizes the exciton with lower energy. The opposite trend is true when the bond length becomes shorter. The above effect gives rise to a negative correlation between pre-edge excitation energies and covalent bond lengths as shown by the red areas and dashed black lines in Fig.~\ref{fig:pre}(a). Moreover, the above negative correlation matches well with the anticorrelation between average covalent bond lengths and pre-edge energies of H$_2$O and D$_2$O as shown by the dashed blue line in Fig.~\ref{fig:pre}(a) which has a negative slope with $\tan\alpha$=0.006 \r{A}/160 meV, where 160 meV is the blue-shift of the pre-edge and 0.006 \r{A} is the bond length contraction when going from H$_2$O (1.005 \r{A}) to D$_2$O (0.999 \r{A}). The $0.6\%$ covalent bond contraction ratio is close to the ratio of $0.5\%$ found in the neutron diffraction experiment of Zeidler {\it{et al.}} \cite{zeidler_oxygen_2011} (the $3\%$ contraction found in the experiment of Soper {\it{et al.}} \cite{soper_quantum_2008} is likely to overestimate the bond contraction as stated in Ref.~\cite{ko_isotope_2019}).
At the same time, pre-edge of D$_2$O has a narrower spectral width than that of H$_2$O as observed in Fig.~\ref{fig:xas}, which is consistent with the suppressed quantum delocalization in D$_2$O as shown in Fig.~\ref{fig:gr}.

Besides the excitation energy, the spectral intensity is also affected by the isotope substitution. According to the selection rule, the transition matrix element is determined by the $p$ character in the excitation \cite{nilsson_x-ray_2010}. Based on symmetry analysis, the intensity of the pre-edge is weak, but not vanishing even for an intact H-bonding environment in crystalline ice \cite{wernet_structure_2004, nilsson_x-ray_2010}. Moreover, the pre-edge intensity is rather sensitive to local distortions of the H-bond. As shown in our analysis in Fig.~\ref{fig:pre}(b), the spectral intensity in pre-edge is largely increased as the water structure is deviated from the ideal tetrahedron by more broken H-bonds, which enhances the $p$ character of quasiparticle exciton. As aforementioned, the NQEs weaken the H-bonding strength compared to classical nuclei. The heavy water is, therefore, less influenced by NQEs due to the heavier nuclei mass as evidenced by the stronger H-bonding environment as displayed in Fig.~\ref{fig:pre}(c). Consistently, a slightly weaker pre-edge intensity of D$_2$O as compared to H$_2$O is seen in both experiment and theory.

\begin{figure*}
	\includegraphics[width=1.0\textwidth]{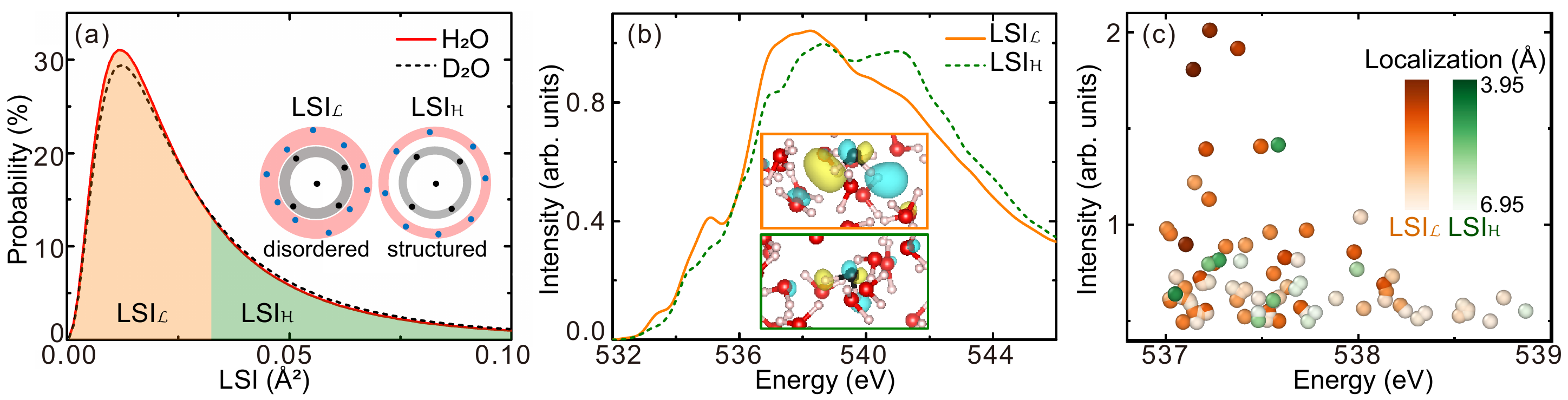}
	\caption{(a) Probability distributions of LSI of liquids H$_2$O (solid red line) and D$_2$O (dashed black line). The yellow and green shadows divide H$_2$O molecules into two parts: LSI$\mathcal{_L}$ and LSI$\mathcal{_H}$. Insets illustrate molecular configurations schematically, with each dot represents a water molecule. The grey and red areas show the first and second coordination shells, respectively.
(b) XAS spectra of liquid H$_2$O contributed by water molecules in part LSI$\mathcal{_L}$ (solid orange line) and part LSI$\mathcal{_H}$ (dashed green line). The XAS spectra of D$_2$O contributed by LSI$\mathcal{_L}$ and LSI$\mathcal{_H}$ molecules have the same trend and are presented in the Supplemental Material \cite{support}. Insets show representative QWs of excited water molecules in part LSI$\mathcal{_L}$ (upper panel) and part LSI$\mathcal{_H}$ (lower panel). (c) Intensities and excitation energies of excitons in the part LSI$\mathcal{_L}$ (yellow circles) and part LSI$\mathcal{_H}$ (green circles). Colors of the circles show the localization distance of QWs, which is defined as the radial distance from the excited oxygen atom that includes $80\%$ of QWs. Excitons with excitation energy $\in$ [537, 539] eV and intensity larger than 0.5, which contribute to the formation of the main-edge, are presented.
}\label{fig:main}
\end{figure*}

The main-edge has been assigned to exciton resonance of $b_2$ characteristic, which originates from the second excited state in a water monomer \cite{chen_x-ray_2010, sun_x-ray_2017}. Because it is unbound, its quasiparticle can no longer be confined within excited water molecules. Nevertheless, main-edge is relatively low in energy, and a certain degree of localization remains. As schematically plotted in the inset of Fig.~\ref{fig:main}(b), quasiparticles of main-edge are largely distributed on the water molecules in first and second coordination shells, which gives rise to notable spectral intensities determining the main-edge feature. As a result, the main-edge serves as a probe of intermediate-range ordering of water \cite{chen_x-ray_2010, sun_x-ray_2017}.

In the intermediate-range, the softer liquid structure in H$_2$O is evidenced by both the less structured first and second coordination shells in the ${g\rm_{OO}}(r)$ [Fig.~\ref{fig:gr}(a)] and the slightly larger density of 0.10623 (0.10007) atom/\r{A}$^3$ of H$_2$O than 0.10581 (0.10000) atom/\r{A}$^3$ of D$_2$O.  In the above, the number outside (within) parentheses denotes the theoretical (experimental \cite{soper_quantum_2008}) values. Therefore, more nonbonded water molecules in H$_2$O will flow into interstitial regions. Indeed, the light water has a slightly less deep first minimum in the ${g\rm_{OO}}(r)$, which is observed in both experiment and theory in Fig.~\ref{fig:gr}(a). With its denser interstitial regions, the light water is more disordered. In order to quantify the degree of inhomogeneity, we resort to the local structure index (LSI) analysis \cite{shiratani_growth_1996, duboue-dijon_characterization_2015, santra_local_2015}. The resulting distributions of LSI are shown in Fig.~\ref{fig:main}(a).
Based on the average LSI value (0.0326 \r{A}$^2$) of H$_2$O, we further decompose the LSI distributions into low LSI (LSI$\mathcal{_L}$) and high LSI (LSI$\mathcal{_H}$) regions in Fig.~\ref{fig:main}(a), which are used to qualitatively describe the disordered and structured configurations, respectively.
As expected, H$_2$O shows itself as a more disordered liquid through the more prominent peak in LSI$\mathcal{_L}$ as compared to D$_2$O. This structural difference is responsible for the observed isotope effect in XAS spectra at the main-edge.

Relative to the structured liquid in LSI$\mathcal{_H}$, the magnitude of disorder increases significantly in LSI$\mathcal{_L}$, whose quasiparticle wavefunctions become more localized in real space simultaneously as shown in Fig.~\ref{fig:main}(c). The disorder-promoted excitation localization is a well-known effect in semiconductors due to the enhanced backscattering processes \cite{klingshirn_semiconductor_2005, tongay_defects_2013}. The similar mechanism applies in water; the surrounding water molecules, similar to defects in semiconductors, serve as the scattering center. Besides, the enhanced localizations also stabilize the excitonic states and give rise to larger transition matrix elements. The above can be clearly seen by the systematically red-shifted energies and much stronger spectral intensities at the main-edge of LSI$\mathcal{_L}$ than those of LSI$\mathcal{_H}$ in Fig.~\ref{fig:main}(b). The light water is composed of a slightly larger fraction of LSI$\mathcal{_L}$, therefore, the resulting main-edge of XAS in H$_2$O is located at lower energy with higher amplitude relative to that of D$_2$O in Fig.~\ref{fig:xas}(a).

The electronic excitations at the post-edge are exciton resonant states as well, which share the same $b_2$ in orbital characteristic \cite{chen_x-ray_2010, sun_x-ray_2017}. However, they are much higher in energy than those in the main-edge. Not surprisingly, the remaining localization in the main-edge is completely absent. At post-edge, the exciton resonances become Bloch-like states that distribute over the entire space of liquids \cite{chen_x-ray_2010, sun_x-ray_2017}. The above delocalized nature makes the post-edge an indicator of the long-range ordering of the H-bond network of water. It can be seen in Fig.~\ref{fig:xas} that the isotope effect at post-edge is negligible, which is consistent with the almost identical ${g\rm_{OO}}(r)$ beyond the second coordination shell \cite{schreck_isotope_2016} [Fig.~\ref{fig:gr} (a)].

\section{CONCLUSION}

In conclusion, we have studied the isotope effect in XAS spectra of water by advanced theoretical methods. The electron-hole excitation was modeled by quasiparticle approach to solving the Bethe-Salpeter equation approximately.
Facilitated by machine learning techniques, the liquid structures were generated from PI-DPMD simulations with the accuracy of the SCAN meta-generalized gradient approximation functional.
Our theoretical simulations have reproduced the isotope effect in XAS spectra of water semi-quantitatively, with the isotopic XAS spectral differences slightly larger than experimental results, which are expected to be improved in future studies with more accurate methods in the description of molecular structure and electron-hole interactions.
The observed blueshifts of spectral energies with weaker intensities, on pre- and main-edge, indicate that the heavy water has a slightly more structured H-bond network in short- and intermediate-range than normal water. This is due to the intricate competing effects from NQEs that affect the heavy water slightly less than normal water. The successful theoretical modeling of the delicate isotope effect on XAS spectra will provide an important means for further exploration of the delicate nature of the H-bond network of water.

\begin{acknowledgments}

This work was supported by National Science Foundation through Awards No. DMR-1552287. This research used resources of the National Energy Research Scientific Computing Center, which is supported by the U.S. Department of Energy (DOE), Office of Science under Contract No. DE-AC02-05CH11231. The work of F. T. and L. Z. (Deep learning molecular dynamics) was supported by the Computational Chemical Center: Chemistry in Solution and at Interfaces funded by the DOE under Award No. DE-SC0019394. This work of J. X. (SCAN-based PI-AIMD) was supported as part of the Center for the Computational Design of Functional Layered Materials, an Energy Frontier Research Center funded by the U.S. Department of Energy, Office of Science, Basic Energy Sciences, under Grant No. DE-SC0012575. This research includes calculations carried out on HPC resources supported in part by the National Science Foundation through major research instrumentation grant number 1625061 and by the US Army Research Laboratory under contract number W911NF-16-2-0189.

\end{acknowledgments}

\bibliography{reference}
\end{document}